%% file: master.tex
\documentclass[12pt,a4paper,titlepage,openany]{report}

\title{Zero-Knowledge Authentication}
\author{Jakob Povšič}

\usepackage{listings}
\usepackage{amsmath}
\usepackage{amsthm}
\usepackage{tgschola}
\usepackage{amsfonts}
\usepackage{mathptmx}
\usepackage{graphicx}
\usepackage[hidelinks]{hyperref}
\usepackage{cite}
\usepackage[english]{babel}
\usepackage[utf8]{inputenc}
\usepackage{anyfontsize}
\usepackage{pgf}
\usepackage{pgfpages}

\usepackage{rotating}

\usepackage{zakljucna_FAMNIT_1_stopnja_MA_MEF_2016}

\normalsize

\usepackage{fancyhdr}

\fontfamily{qcs}

\newcommand{\Mod}[1]{\ (\mathrm{mod}\ #1)}

\newcommand{\genlegendre}[4]{%
  \genfrac{(}{)}{}{#1}{#3}{#4}%
  \if\relax\detokenize{#2}\relax\else_{\!#2}\fi
}
\newcommand{\legendre}[3][]{\genlegendre{}{#1}{#2}{#3}}
\newcommand{\dlegendre}[3][]{\genlegendre{0}{#1}{#2}{#3}}

\newtheorem*{remark}{Definition}

\theoremstyle{definition}

\begin{document}
\newpage
\input{keywords}

\newpage

\pagenumbering{arabic}

\input{chapters/introduction/introduction}

\input{chapters/methodologies/introduction}
\input{chapters/methodologies/authentication}
\input{chapters/methodologies/password-authentication}
\input{chapters/methodologies/eap}

\input{chapters/methodologies/interactive-zkps}

\input{chapters/methodologies/languages-with-zkps}

\input{chapters/results/intro}

\input{chapters/results/pba-using-qr-zkp}
\input{chapters/results/eap_pzkpqr}
\input{chapters/results/conclusion}

\input{chapters/slo-summary}

\fancypagestyle{plain}{}
\bibliographystyle{plain}
\bibliography{ref}

\end{document}

%% file: keywords.tex
\pagestyle{empty}
\reversemarginpar
\marginpar{
\vspace{-1cm}
\hspace{-2cm}
\begin{sideways}
\large POVŠIČ \hspace{2cm} ZAKLJUČNA NALOGA (FINAL PROJECT PAPAER) \hspace{8cm} 2022
\end{sideways}
}

\begin{center}
\vspace{-1cm}
\noindent \large UNIVERZA NA PRIMORSKEM\\
\large FAKULTETA ZA MATEMATIKO, NARAVOSLOVJE IN\\
INFORMACIJSKE TEHNOLOGIJE

\vspace*{\fill}
\large ZAKLJUČNA NALOGA\\
\large (FINAL PROJECT PAPER)\\
\vspace{0.3cm}
\textbf{\Large AVTENTIKACIJA Z DOKAZI NIČELNEGA ZNANJA}\\
\textbf{\Large (ZERO-KNOWLEDGE AUTHENTICATION)}
\vspace*{\fill}
\vspace{1.8cm}
\end{center}

\begin{flushright}
\noindent \large JAKOB POVŠIČ\\
\vspace{2cm}
\end{flushright}

\newpage

\pagestyle{empty}
\begin{center}
\noindent \large UNIVERZA NA PRIMORSKEM\\
\large FAKULTETA ZA MATEMATIKO, NARAVOSLOVJE IN\\
INFORMACIJSKE TEHNOLOGIJE

\normalsize
\vspace{6cm}
Zaključna naloga\\
(Final Project Paper)\\
\textbf{\large Avtentikacija z dokazi ničelnega znanja}\\
\normalsize
(Zero-Knowledge Authentication)\\
\end{center}

\begin{flushleft}
\vspace{5cm}
\noindent Ime in priimek: Jakob Povšič
\\
\noindent Študijski program: Računalništvo in informatika
\\
\noindent Mentor: prof. dr. Andrej Brodnik
\\
\end{flushleft}

\vspace{4cm}
\begin{center}
\large \textbf{Koper, maj 2022}
\end{center}
\pagenumbering{Roman}
\newpage
\pagestyle{fancy}

\section*{Ključna dokumentacijska informacija}

\medskip
\begin{center}
\fbox{\parbox{\linewidth}{
\vspace{0.2cm}
\noindent
Ime in PRIIMEK:\vspace{0.5cm} Jakob POVŠIČ\\
Naslov zaključne naloge:\vspace{0.5cm} Avtentikacija z ničelnega dokazi\\
Kraj:\vspace{0.5cm} Koper\\
Leto:\vspace{0.5cm} 2022\\
Število listov: 46\hspace{2cm} Število slik: 6\hspace{2.6cm} Število tabel: 12\hspace{2cm}\vspace{0.5cm}\\
\hspace{1cm} Število referenc: 57\vspace{0.5cm}\\
Mentor:\vspace{0.5cm} prof. dr. Andrej Brodnik\\
Ključne besede: razširljivi aventikacijski protokol (EAP), dokazi ničelnega znanja, avtentikacija, kriptografija, raztegovanje ključev, gesla, problem kvadratnih ostankov\vspace{0.5cm}\\
Izvleček:\\
V zaključnem delu se osredotočamo na zasnovo sistema za aventikacijo uporabnikov prek omreżja z uporabniskim imenom in geslom. Sistem uporablja dokaze ničelnega znanja (ZKP) kot mehanizem preverjanja gesla. Uporabljen ZKP protokol je zasnovan na problemu kvadratnih ostankov. Avtentikacijski sistem je oblikovan kot metoda v razširljivem aventikacijskem protokolu (EAP). Uporaba ZKP sistema nam poda varnostne lastnosti, ki so primerne za uporabno na nezaščitenih omrezjih.
\vspace{0.2cm}
}}
\end{center}

\newpage

\section*{Key words documentation}

\medskip

\begin{center}
\fbox{\parbox{\linewidth}{
\vspace{0.2cm}
\noindent
Name and SURNAME: Jakob POVŠIČ\vspace{0.5cm}\\
Title of final project paper:\vspace{0.5cm} Zero-Knowledge Authentication\\
Place: Koper\vspace{0.5cm}\\
Year:2022\vspace{0.5cm}\\
Number of pages: 46\hspace{1.6cm} Number of figures: 6\hspace{2.2cm}Number of tables: 12\vspace{0.5cm}\\
\hspace{0.8cm}Number of references: 57\vspace{0.5cm}\\
Mentor: Prof. Andrej Brodnik, PhD\vspace{0.5cm}\\
Keywords: Extensible Authentication Protocol, Zero-Knowledge Proofs, Authentication, Cryptography, Key-Stretching, Passwords Authentication, Quadratic Residuosity Problem\vspace{0.5cm}\\
Abstract:\\In the thesis we focus on designing an authentication system to authenticate users over a network with a username and a password.
The system uses the zero-knowledge proof (ZKP) system as a password verification mechanism.
The ZKP protocol used is based on the quadratic residuosity problem.
The authentication system is defined as a method in the extensible authentication protocol (EAP).
Using a ZKP system yields interesting security properties that make the system favourable to be used over insecure networks.
\vspace{0.2cm}
}}
\end{center}
\newpage
\section*{Acknowledgement}
I would like to express my gratitude towards Prof. Andrej Brodnik for his guidance, patience, and persistence, all of which were essential in making of this work.
I would also like to thank my family and friends for supporting and believing in my efforts, and to my work colleagues for entrusting me with the responsibility of solving hard problems.

\newpage

\tableofcontents
\addtocontents{toc}{\protect\thispagestyle{fancy}}
\newpage
\listoftables
\addtocontents{lot}{\protect\thispagestyle{fancy}}
\newpage
\listoffigures
\addtocontents{lof}{\protect\thispagestyle{fancy}}
\newpage
\renewcommand{\cftdot}{}
\thispagestyle{fancy}
\newpage

\chapter*{List of Abbreviations}
\thispagestyle{fancyplain}
\begin{longtable}{@{}p{1cm}@{}p{\dimexpr\textwidth-1cm\relax}@{}}
\nomenclature{$ZKP$}{Zero-Knowledge Proof}
\nomenclature{$EAP$}{Extensible Authentication Protocol}
\end{longtable}
\newpage

\normalsize

%% file: chapters/introduction/introduction.tex
\newpage
\chapter{Introduction}
\thispagestyle{fancy}
\label{chapter:1}

\noindent
Today privacy is a necessary sacrifice we have to make in order to take part in the digital world, imperative to our modern life.
Every day, more digital systems gain access to our personal information. While this practice is often a necessary evil, many companies seek to exploit this position.
Zero-knowledge proofs (ZKPs) have the potential to change how our data exists in the digital space. 
ZKP systems are an intriguing cryptographic phenomenon for proving mathematical statements without revealing \textit{why} they are true.

Cryptocurrencies like Zcash \cite{hopwood2016zcash} use ZKPs to confirm transactions while keeping the sender and the recipient anonymous and the transaction amount opaque.
The self-sovereign identity space \cite{tobin2016inevitable} uses ZKPs as an essential part in a decentralized and privacy-preserving digital identity infrastructure.
ZKPs enable a system to verify the properties \cite{10.1007/978-3-540-89255-7_15} of sensitive data without seeing it and risking misuse. With these tools, we can prove legal age, financial solvency, or our nationality, while revealing no sensitive information.

Advances like ZKPs hint of a future where we will look at our current personal data practice as feudal and undignified.

\bigskip
\noindent
The focus of this thesis will be to define a simple use for ZKPs.
We will design an authentication system using a ZKP as a password verification method, as an authentication method in the extensible authentication protocol (EAP).
Moreover, in the system design we also have to consider standard methods for protecting against inherent vulnerabilities of password based systems.
\section{Structure of the thesis.}
This thesis is composed of three chapters.
In \S\ref{chapter:2} we explore the two key topics of authentication and ZKPs.
In \S\ref{chapter:3} we present the architecture of our authentication system and the extensible authentication protocol method definition.
Chapter \S\ref{chapter:4} concludes the thesis and gives some possible ways for future work.

\newpage

%% file: chapters/methodologies/introduction.tex
\chapter{Methodologies and Tools}
\thispagestyle{fancy}
\label{chapter:2}

\noindent
In this chapter, we will explore the concepts and components behind our authentication system.
To design a password authentication system, we must first understand what an authentication system is, the mechanism of password based authentication, and how to avoid its vulnerabilities.
We will also look into the EAP framework, into which we will build our authentication system.
Our system will use a ZKP as a password verification mechanism, so we also examine what ZKPs are, how they work, and the mathematics behind why do they work.

%% file: chapters/methodologies/authentication.tex
\section{Authentication}

Authentication is the process of proving a claim or an assertion.
Today, the term is most commonly used in information security \cite{shirey2007internet}, however, we can find the principles of authentication in fields of archeology, anthropology and others \cite{Odegaard2014}.

In computer science, we commonly use authentication for establishing access rights between protected system resources and users through digital identities.
Government and international institutions have developed guidelines for managing digital identities and authentication processes \cite{grassi2017}.

While systems can authenticate both humans and other computer systems, we are focusing on authentication of a human end-user.

\subsection{Authentication Process Components}
Authentication \cite{shirey2007internet} is verifying a claim that an entity or a resource has a certain attribute value.
This is a broad definition, and it most frequently applies to the verification of a user's identity (e.g. at login), however we can make and verify claims about any subject or object.
The process of authentication is done in two parts, \textit{identification} and \textit{verification}.
A common application of authentication is to manage access of an external user to protected system resources.

\paragraph{Identification.} Presenting an identifier to the authentication system that establishes the entity being authenticated. This is commonly a username or an email address. The identifier needs to be unique for the entity it identifies.

The process of identification is not necessarily externally visible, as the identity of the subject can be implicit in the environment. 
For example, we can determine an identifier from the user’s IP address.
Or a system might only have a single user that needs to be authenticated.

\paragraph{Verification.} Presenting or generating authentication information that can verify the claim.
Commonly used authentication information are passwords, one time tokens, digital signatures, etc.

Our system will verify the user's password using a ZKP.

\subsection{Authentication Factors}

Authentication systems can rely on three groups of factors \cite{bignell2006authentication}.

\begin{description}
	\item [Knowledge factors] Something the user \textbf{knows} (e.g. password, security question, PIN)
	\item [Ownership factors] Something the user \textbf{owns} (e.g. ID card, security tokens, mobile devices)
	\item [Inherence factors] Something the user \textbf{is} or \textbf{does} (e.g. static biometrics - fingerprints, retina, face. dynamic biometrics - voice patterns, typing rhythm)
\end{description}

\paragraph{Strong authentication.}
As defined by governments and financial institutions \cite{cnss2006national, ecb2013recommendations}, strong authentication is a system using two or more factors. 
We also referred to this as \textit{multi-factor authentication}. Our system will focus only on the user possessing a password (\textit{knowledge factor}), while the relying party can use additional authentication factors to improve security.

%% file: chapters/methodologies/password-authentication.tex
\section{Password Authentication}
\label{section:password-authentication}

Passwords are one of the most common and oldest forms of user authentication, being first used in computers at MIT in the mid-60s \cite{mcmillan2012password}.

Let us examine the high-level model of password authentication, its risks and tools to mitigate them.

\subsection{Authentication Model}

\begin{figure}[h]
	\centering
	\includegraphics[height=8cm]{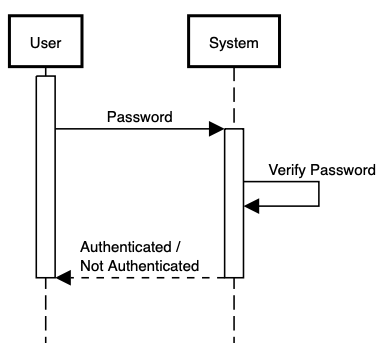}
	\caption{Password Authentication Model}
	\label{fig:password-authentication}
\end{figure}

Password authentication is a simple model (Figure  \ref{fig:password-authentication}) based on a shared secret between the user and the system. The secret is usually a set of characters or words memorised by the user, inputted via a keyboard.
We often used the password in a combination with a username.
To authenticate, the user simply needs to prove to the system his knowledge of the password.

\subsection{Security Vulnerabilities and Attacks}
\label{label:password-vulnerabilities}

In a common password authentication system used on the web, the user sends a plain-text password over a secure HTTPS connection, the server verifies it and responds.
The simplicity that makes passwords practical for users makes them vulnerable.

Authors \cite{conklin2004password} have shown that users pick passwords that are easier to remember and reuse the same passwords across different websites.
Many websites also don't properly handle passwords, permitting attackers to access plain-text passwords when a security breach happens.
These vulnerabilities can result in mild inconveniences to serious offences like identity theft.
The industry is aiming to improve password security with the adoption of password managers and initiatives like FIDO \cite{cho2014passwordless} working to retire passwords altogether.

The National Institute of Standards and Technology \cite{grassi2017} classifies attacks as \textit{online} or \textit{offline}, based on how the attacker is interacting with the system.

\subsubsection{Online Attacks}
An attack where an attacker is directly interacting with the authentication system.
These attacks are usually very \textit{noisy}, making it easy for an authentication system to detect and react.
For example, locking an account after a few failed authentication attempts.
For this reason, brute-force online attacks are not effective.

Effective online attacks use the strategy of appearing as normal users, thus remaining under the radar of detection.
Popular methods are \textit{password spraying} and \textit{credential stuffing} \cite{haber2020attack}, both of which utilise information from data breaches, like lists of most commonly used passwords, or username and password combinations.
Password spraying is taking a few commonly used passwords and attempting to authenticate with many accounts. The attacker is assuming that in a large sample of users some will use these weak passwords.
Credential stuffing is taking a compromised user credential, for example, a username and password combination found in a data breach, and using it to authenticate into multiple websites.
The attacker is assuming that if a person is using a set of credentials on one website, they are potentially reusing them on others.

\subsubsection{Offline Attacks}
Are attacks performed in a system controlled by the attacker.
For example, an attacker might analyse data on his personal computer to extract sensitive information.
The data is obtained by either theft of a file, eavesdropping on an authentication protocol or a system penetration.

\textit{Password cracking} \cite{blocki2018economics} is a method of extracting passwords from data used by the authentication system for password verification.
Two parameters determine the chance of success when password cracking, the time required to check a single password and number of guesses required or the strength of the underlying password.

\subsubsection{Security Practices}
\label{password-security-practices}
There are many practices an authentication system can incorporate to improve its security.
The authentication system uses the persistently stored sensitive data for password verification. This data could be used by an attacker in an offline attack, so we need to protect them somehow.
Often the password verification method imposes constraints on how we can transform the sensitive password verification data. Later, we will examine which constraints our ZKP system imposes.
We are going to focus on methods for improving the security of the data layer against offline attacks.
A production ready system should adopt many other security practices, however this is outside the scope of this thesis.

\paragraph{Key-Stretching}
\label{paragraph:password-hashing}
Protecting password verification data in persistent storage is of critical importance.
\textit{Key-stretching} \cite{hornby2016salted} also called \textit{password hashing} is the industry standard method of improving security of low entropy secrets like passwords.

With this approach the password $p$ is \textit{stretched} or \textit{hashed} using a function $H$ and a high entropy value called a \textit{salt} $s$, $H(p, s) = p_H$. The output called a \textit{password hash} $p_H$ and the salt are persistently stored while the plain text password is discarded. When verifying the password $p'$, it is stretched again $H(p', s) = p{'}_H$ with the stored salt $s$ and the output hash $p{'}_H$ is compared with the stored password hash $p{'}_H \stackrel{?}{=} p_H$, if it matches the password is correct.

Key-stretching \cite{blocki2018economics} is traditionally done with CPU intensive hash iteration functions (PBKDF2, Bcrypt) \cite{kaliski2000pkcs, provos1999bcrypt}. Recently, however memory-hard algorithms (Argon2, Scrypt, Balloon) \cite{biryukov2016argon2, percival2016scrypt, boneh2016balloon} are becoming a standard choice, because they protect against using special purpose hardware (ASIC) for calculating hashes.

%% file: chapters/methodologies/eap.tex
\section{Extensible Authentication Protocol}
\label{section:eap}
We will define our authentication system as a method in the extensible authentication protocol (EAP) framework.

Extensible authentication protocol \cite{aboba2004extensible} (EAP) is a general purpose authentication framework designed for network access authentication. 
It runs directly over the data link layer, such as PPP \cite{simpson1994rfc1661} and IEEE 802 \cite{10.5555/18422.18423}.
EAP defines a set of messages that support negotiation and execution of a variety of authentication protocols.
EAP is a two-party protocol between a \textit{peer} and an \textit{authenticator} at each end of a link.  

\subsection{Messages}
\begin{figure}[h]
	\centering
	\includegraphics[width=8cm]{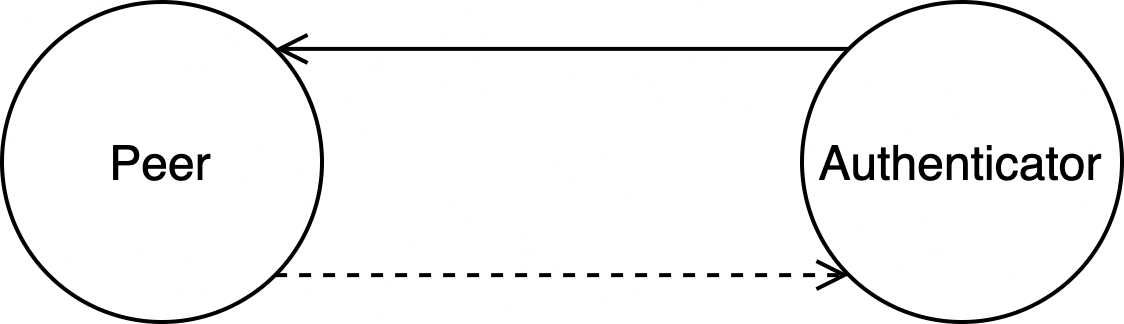}
	\caption{EAP Peer and Authenticator Communication}
	\label{fig:eap-messages}
\end{figure}

The peer and the authenticator communicate by exchanging \textit{EAP messages} (Figure \ref{fig:eap-messages}).
The protocol starts with the authenticator sending a message to the peer. They keep exchanging messages until the authenticator can either authenticate the peer or not.

Messages are exchanged in a lock-step manner, where an authenticator sends a message and the peer responds to it. 
The authenticator dictates the order of messages, meaning it can send a message at any point of communication, as opposed to the peer, which can only respond to messages from the authenticator.
Any messages from the peer not in a response to the authenticator are discarded.
\bigskip

\subsubsection{Message Structure}
Messages are composed of fields (Table \ref{table:eap-message}), each field length is multiple of an octet of bits.
Each field has a special purpose:

\begin{table}
	\caption{EAP Message Structure}
	\vspace{0.1cm}
	\centering
	\begin{tabular}{|c|c|c|c|c|c|}
		\hline
		Length (Octets) & 1 & 1 & 2 & 1 & $n \le 2^{16}$\\
		\hline
		Field & Code & Identifier & Length & Type & Type Data\\
		\hline
	\end{tabular}
	\label{table:eap-message}
\end{table}

\paragraph{Code Field}
Determines who the packet is intended for and how or even should the recipient respond.
The following code values are defined:

\begin{quote}
\begin{description}
	\item[Request]\textit{Code 1}. Messages sent by the authenticator to the peer. Response is always expected.
	\item[Response]\textit{Code 2}. Messages sent by the peer to the authenticator as a reply to a \textit{request} message.
	\item[Success]\textit{Code 3}. Sent by the authenticator, when the peer is successfully authenticated. The peer doesn't respond to the message.
	\item[Failure]\textit{Code 4}. Sent by the authenticator, when the peer cannot be authenticated. The peer doesn't respond to the message.
\end{description}
\end{quote}

\paragraph{Identifier Field}
Used to create a session between the peer and the authenticator.
The authenticator uses the field to match request and response messages, each response message needs to have the same identifier as the request.
The authenticator will discard response messages that don't have a matching identifier with the current request.
The peer does not re-transmit a response message, but relies on the authenticator to re-transmit a request message after some time if the matching response is lost.

\paragraph{Length Field}
Determines the total size of the EAP message. Because EAP provides support for generic authentication methods, the final length of the messages is variable.
The length of the \textit{type data} field entirely depends on the authentication method used.

\paragraph{\textit{Type} and \textit{Type Data} Field}
\label{text:eap:type}
Determines how the message should be processed and how to interpret the \textit{type data} field.
Most message types represent authentication methods, except for four special purpose types.
The \textit{type} used is determined by the authenticator when sending the request message. The response message from a peer needs to be of the same type as the request, except where that type is not supported by the peer.
The following types are defined:

\begin{quote}
\begin{description}
	\label{def:eap-identitiy}
	\item[Identity] \textit{Type 1}. Used to query the identity of the peer. The type is often used as an initial message from the authenticator the peer, however its use is entirely optional and EAP methods should rely on method-specific identity queries.
	
	\item[Notification]\textit{Type 2}. Used to convey an informative message to the peer, by the authenticator. Usage of this type is entirely optional.
	\item[Nak]\textit{Type 3}. Used only as a response to a request, where the desired type is not available.
	The peer includes desired authentication methods, indicated by their type number.
	This type is also referred to as Legacy Nak, when compared to \textit{Expanded Nak} (sub-type of the Expanded Type).
	\item[Expanded Type] \textit{Type 254}. 
	Used to expand the space of possible message types beyond the original 256 possible types.
	The expanded type \textit{data field} is composed from a \textit{Vendor-ID} field, \textit{Vendor-Type} and the type data.
	\bigskip
	\begin{center}
		\begin{tabular}{|c|c|c|c|c|c|}
		\cline{2-5}
		\hline
		Length (octets) & & 3 & 4 & n\\
		\hline
		Field Type & ... & Vendor-ID & Vendor-Type & Vendor-Type Data\\
		\hline
		\end{tabular}
	\end{center}
	\bigskip
	A peer can respond to an unsupported request type with an \textit{expanded nak}, if he desires to use an EAP method supported with the expanded type.
	\item[Experimental] \textit{Type 255}. This type is used for experimenting with new EAP Types and has not fixed format.
\end{description}
\end{quote}

\paragraph{Authentication Methods}
The remaining types correspond to different authentication methods.
IANA \cite{joseph2004eap} assigns type numbers to 49 different authentication methods.
Authors of the original RFC \cite{aboba2004extensible} already assigned 3 authentication protocols:

\begin{quote}
\begin{description}
	\item[MD5-Challenge] \textit{Type 4}. An EAP implementation of the PPP-CHAP protocol \cite{simon2008eap}.
	\item[One-Time Password] \textit{Type 5}. An EAP implementation of the one-time password system \cite{haller1998one}.
	\item[Generic Token Card] \textit{Type 6.} This type facilitates various challenge/response \textit{token card} implementations.
\end{description}
\end{quote}

Some other notable examples are EAP-TLS \cite{simon2008eap}, EAP-PSK \cite{bersani2007eap}.
EAP SRP-SHA1 \cite{ietf-pppext-eap-srp-03} is especially interesting as it uses a ZKP system to verify the peer's secret, similar to our own EAP method.

\subsection{Pass-Through Behaviour}
An authenticator can act as a \textit{Pass-Through Authenticator}, by using the authentication services of a \textit{backend authentication server}.
In this mode of operation, the authenticator is relaying the EAP messages between the peer and the backend authentication server.
For example, in IEEE 802.1x the authenticator communicates with a RADIUS server \cite{congdon2003ieee}.

\paragraph{IEEE 802.1x}

Is a port based network access control standard for LAN and WLAN.
It is part of the IEEE 802.11 group of network protocols.
IEEE 802.1x defines an encapsulation of EAP for use over IEEE 802 as EAPOL or "EAP over LANs".
EAPOL is used in widely adopted wireless network security standards WPA2. 
In WPA2-Enterprise, EAPOL is used for communication between the supplicant and the authenticator.

With WPA2-Enterprise, the authenticator functions in a pass-through mode and uses a RADIUS server to authenticate the supplicant.
EAP packets between the authenticator and the authentications server (RADIUS) are encapsulated as RADIUS messages \cite{aboba2003radius, chen2005extensible, congdon2003ieee}.

\subsection{MD5-Challenge EAP Method}
Let us look at an example of an EAP authentication process and examine the messages exchanged between the peer and the authenticator. This EAP instance uses the MD5-Challenge authentication method.
MD5-Challenge is an EAP method analogous to PPP CHAP \cite{simpson1996ppp}. The message Type 4 denotes the method.

\paragraph{PPP CHAP.}
The PPP \textit{challenge handshake authentication protocol} is an authentication model based on a shared secret between the peer and the authenticator.
The authenticator authenticates the user by first sending him a random challenge $c$, which the user concatenates with the secret $s$ and hashes with a hashing function $d = h(c | s)$.
The hash digest $d$ is returned in the response and the authenticator compares the received hash with the locally computed hash, if they match the peer is authenticated.

Aside from a slightly different message format, the MD5-Challenge authentication process is functionally the same as PPP CHAP, with the difference that while PPP CHAP is hashing algorithm agnostic, MD5-Challenge specifics the use of the MD5 hashing algorithm \cite{rivest1992md5}.

\begin{figure}[h]
	\centering
	\includegraphics[width=10cm]{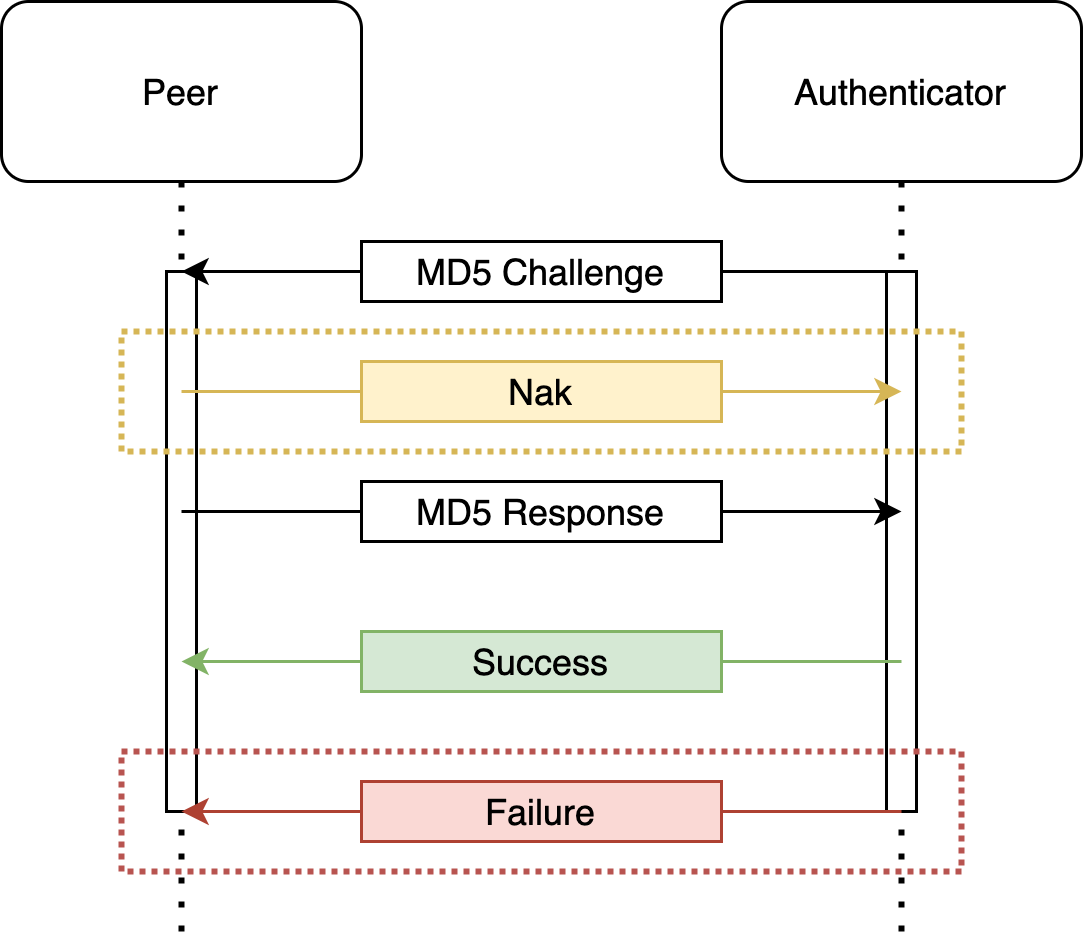}
	\caption{EAP (MD5-Challenge)}
	\label{fig:eap-md5}
\end{figure}

\bigskip
\noindent
Let us examine the individual steps in EAP authentication with this method. 
The steps are visualised in the sequence diagram \ref{fig:eap-md5}.
At each step we will describe what is happening and note the contents of the EAP messages being exchanged, we are omitting the \textit{identifier}, \textit{length} and \textit{type-data} fields of the messages, as their contents are dynamically determined when the protocol is running.

\begin{description}
	\item [MD5-Challenge] Message is sent to the peer with a random challenge $c$. $(Code=1, Type=4)$.
	\item [Nak] In the case that MD5-Challenge method is unacceptable to the peer, he should respond with a \textit{Nak} message $(Code=2, Type=3)$, the \textit{type data} of the message can contain the number indicating the preferred EAP method.
	\item [MD5-Response] The peer computes the hash digest $d$ as described before with the MD5 hashing algorithm and sends it in a response. $(Code=2, Type=4)$.
	\item [Success] The authenticator sends this message if the digest $d$ was successfully verified. The peer was successfully authenticated, and the protocol stops. $(Code=3)$.
	\item [Failure] If the digest $d$ isn't valid, the authenticator sends the \textit{failure} message indicating that the peer could not be authenticated. $(Code=4)$.
\end{description}

%% file: chapters/methodologies/interactive-zkps.tex
\section{Zero-Knowledge Proofs}
In our authentication system, we wish to use a ZKP as a password verification method.

In this section we explore what ZKPs are on a high level, look at a practical analogy of how they work and also how they are used in real life.
Next we look at what are \textit{interactive proof systems}, the parent of ZKP systems. How to quantify the knowledge exchanged in an interactive proof system and finally what makes an interactive proof system zero-knowledge.

\subsection{Basics}
ZKPs are a concept for proving the validity of mathematical statements.
What makes them particularly interesting is that ZKPs can prove a statement revealing no information about why a statement is true, hence the term \textit{zero-knowledge}.

In mathematics, theorems are logical arguments that establish truth through inference rules of a deductive system based on axioms and other proven theorems.
ZKPs are probabilistic, meaning they \textit{convince} the verifier of the validity with a negligible margin of error.
We use the term convince, because ZKPs are not absolute truth, but the chance of a false statement convincing a verifier is arbitrarily small. 
The difference in definition is subtle, but we will see what that means in practice further on.

ZKPs were first described by Goldwasser, Micali and Rackoff in \cite{goldwasser1989knowledge} in 1985. 
They proposed a proof system as a two-party protocol between a \textit{prover} and a \textit{verifier}.

To help our understanding, we will explore the strange cave of Ali Baba, a famous analogy for a zero-knowledge protocol from a publication called “How to explain zero-knowledge protocols to your children” \cite{10.1007/0-387-34805-0_60}.

\subsubsection{The Strange Cave of Ali Baba}

\bigskip

Ali Baba's cave has a single entrance that splits into two tunnels that meet in the middle, where there is a door that only opens with a secret passphrase.

\begin{figure}[h]
	\centering
	\includegraphics[height=6cm]{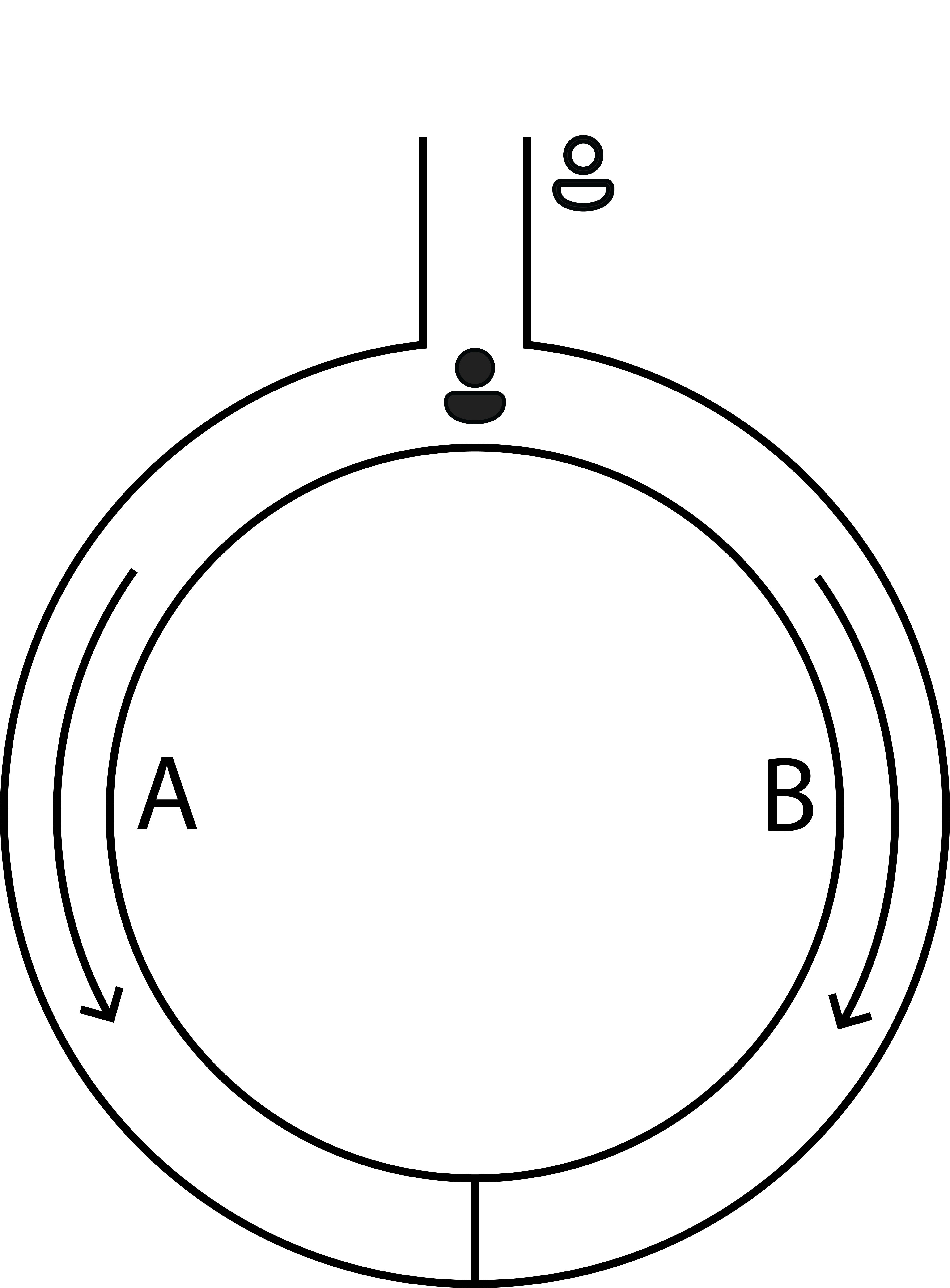}
	\caption{The Strange Cave of Ali Baba}
	\label{fig:strange-cave-of-alibaba}
\end{figure}

\bigskip

Peggy (or Prover) wants to prove to Victor (or Verifier) that she knows the passphrase, but she doesn't want to reveal it nor does she want to reveal her knowledge of it to anyone else besides Victor.

\bigskip

To accomplish this, they come up with a scheme.
Victor stands in front of the cave and faces away from the entrance, to not see Peggy as she enters the cave, and goes into one tunnel at random.
Victor looks at the entrance, so he can see both tunnels, and signals Peggy which tunnel to come out from.
Peggy, knowing the passphrase, can go through the door in the middle and emerge from the tunnel requested.

\bigskip

If Peggy did not know the secret, she could fool Victor, only by entering the correct tunnel by chance.
But since Victor is choosing the tunnel at random, Peggy's chance of picking the correct tunnel is $\frac{1}{2}$. If Victor was to repeat the process $m$ time, her chances of Peggy fooling him become arbitrarily small ($\frac{1}{2^m}$).

This way Peggy can convince Victor that she knows the secret with an arbitrarily high probability of ($1 - \frac{1}{2^m}$).

\bigskip

Any third party observing the interaction cannot be convinced of the validity of the proof because they cannot be assured that the interaction was truly random. 
For example, Victor could have told Peggy his questions in advance, so Peggy would produce a convincing looking proof.

\subsubsection{Applications}
There have been several applications in the blockchains and decentralised identity systems.
The cryptocurrency Zcash uses a \textit{non-interactive zero-knowledge protocol} zk-SNARK \cite{bowe2018multi} to prove the validity of transactions, revealing nothing about the recipients or the amount sent.
Alternativley, \textit{Idemix} \cite{camenisch2002design} an anonymous credential system for interaction between digital identities relies on CL-signatures \cite{camenisch2001efficient} to prove validity of a credential offline, without the issuing organisation.
Idemix has been implemented in the open-source Hyperledger projects.

ZKPs can also prove that values satisfy complex constraints like range proofs \cite{camenisch2008efficient}.

\subsection{Interactive Proof Systems}
\label{section:interactive-proof-systems}
An interactive proof system is a proof system where a \textit{prover} attempts to convince a \textit{verifier} that a statement is true.
The prover and the verifier interact with each other by exchanging data until the verifier is convinced or not.

\begin{figure}[h]
	\centering
	\includegraphics[width=8cm]{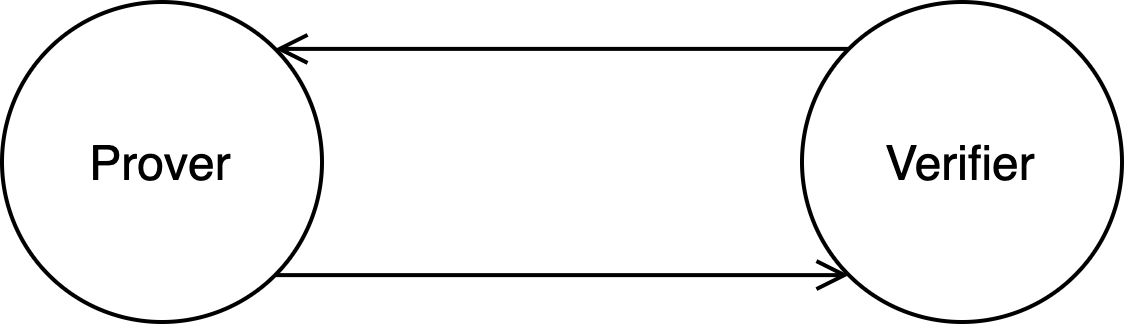}
	\caption{Interactive Proof System}
	\label{fig:interactive-proof-system}
\end{figure}

The prover is a computationally unbounded polynomial time Turing machine and the verifier is a probabilistic polynomial time Turing machine.
An interactive proof system is defined by properties \textit{completeness} and \textit{soundness}.
\newpage
\paragraph{Notation}
\begin{description}
	\item $\Pr[A]$: probability of event $A$ happening.
	\item $P(x) = y$: prover $P$, outputs a proof $y$ for statement $x$.
	\item $V(y) = 1$: verifier $V$, verifies proof $y$ and outputs $1$ or $0$.
	\item $L$: Language $L$ is a set of \textit{words} or values $x$, with a property $P$, $L = \{x \mid P(x)\}$. Often used to described a set of values that are a solutions to a mathematical problem. For example, a problem, find $x$ larger than $3$ and lesser than $7$, has the associated language $L_A$ is $L_A = \{x \mid 3<x<7\} = \{4,5,6\}$. Proving membership of $x$ in $L$ is equivalent to proving $x$ is a solution to a problem $P$ that defines the language $L$.
\end{description}

\paragraph{Completeness}

Any honest prover can convince the verifier with overwhelming probability.\\
For $x \in L$ and each $k \in \mathbb{N}$ and sufficiently large $n$;

$$\Pr[x \in L; P(x) = y; V(y) = 1] \ge 1 - \frac{1}{n^k}$$

\paragraph{Soundness}

Any verifier following the protocol will reject a cheating prover with overwhelming probability.\\
For $x \notin L$ and each $k \in \mathbb{N}$ and sufficiently large $n$;

$$\Pr[x \notin L; P(x) = y; V(y) = 0] \ge 1 - \frac{1}{n^k}$$

\subsection{Zero-Knowledge}

ZKP systems prove the membership of $x$ in language $L$, revealing no additional knowledge (e.g. why is $x \in L$).\\
The essence of zero-knowledge is the idea that the data the verifier has (from current and past interactions with the prover) is indistinguishable from data that can be simulated from public information.
For example, if we return to our analogy in the introduction. 
Victor wants to record what he sees to analyse, or to prove to someone else that Peggy knows the secret.
Victor records which tunnels he calls and from which Peggy emerges, he doesn't record which tunnel Peggy goes into as he is facing away.
Later on Bill and Monica decide to record a similar scheme without knowing the secret.
Bill records himself calling the tunnels and Monica emerging randomly. Sometimes she emerges from the correct one, other times she doesn't. 
Bill later edits the video to only show the times Monica correctly emerged from the tunnel, as if she knew the secret.
Assuming Bill's video editing skills are good, the videos Bill and Victor recorded are indistinguishable, both videos feature someone calling tunnels and a person correctly emerging. 
While Victor's video recorded a genuine proof, there is no information in the video from where we could prove that.
The only one who can be truly convinced is Victor, because he trusts that his own choices of tunnels to call were truly random.

\subsubsection{Indistinguishability}
Indistinguishability describes the (in)ability of distinguishing between two sets of data. The \textit{data} we are comparing is formalised as a random variable.
\bigskip
\newline
Let $U$ and $V$ be two families of random variables and $x \in L$.
We are given a random sample $x$ from either distribution $U$ or $V$, we study the sample to learn which distribution was the origin of $x$.
$U$ and $V$ are said to be \textit{indistinguishable} when our studying of $x$ is no better than guessing randomly.

\subsubsection{Approximability}

The notion of approximability described the degree to which a process $M$ could \textit{generate} a distribution $M(x)$ that is indistinguishable from some distribution $U(x)$.

Formally, a random variable $U(x)$ is \textit{approximable} if there exists a probabilistic Turing machine $M$, such that for $x \in L$, $M(x)$ is \textit{indistinguishable} from $U(x)$.

\subsubsection{Definition of Zero-Knowledge}
Zero-knowledge is a type of interactive proof system in which we can’t learn any meaningful information, besides the validity of the proof.
\bigskip
\newline
An interactive proof system is \textit{zero-knowledge} if $V(x)$ data available to the verifier is \textit{approximable} by $S(x)$ data that can be generated by a \textit{simulator} $S$ from public information.
This also accounts for additional data that might be available to a cheating verifier, for example past interactions with the prover.

\subsubsection{Strengths of Zero-Knowledge}
There are three levels of zero-knowledge, defined by the strength of indistinguishability.
We have defined \textit{indistinguishability} as the ability of a \textit{judge} to distinguish between random variables $V(x)$ and $S(x)$, by attempting to determine the origin of a sample $x$, taken randomly from either distribution.
The strength of indistinguishability is determined by two parameters, the available \textit{time} to analyse and the \textit{size} of the sample.\\
\newline
$V(x)$ represents the verifiers view and $S(x)$ the generated data by the simulator $S$. Or if we return to our earlier analogy, $V(x)$ represents Victors interaction with Peggy, and $S(x)$ a fabricated recording of an interaction between Bill and Monica.

\begin{description}
	\item[Perfect Zero-Knowledge]
	$V(x)$ and $S(x)$ are \textbf{equal} when they remain indistinguishable even when given arbitrary time and an unbounded sample size.
	
	\item[Statistical Zero-Knowledge] Two random variables are \textbf{statistically indistinguishable} when they remain indistinguishable given arbitrary time and a polynomial sized sample.
	\bigskip
	\\
	Let $L \subset \{0,1\}^*$ be a language. Two polynomial sized families of random variables $V$ and $S$ are \textit{statistically indistinguishable} when,
		\medskip
	$$\sum_{\alpha \in \{0,1\}^*} |P[V(x) = \alpha] - P[S(x) = \alpha]| < |x|^{-c}$$
	for all constants $c>0$ and all sufficiently long $x \in L$.
	
	\item[Computational Zero-Knowledge] $V(x)$ and $S(x)$ are \textbf{computationally indistinguishable} when they remain indistinguishable given polynomial time and a polynomial sized sample.
	\bigskip
	\\
	Let $L \subset \{0,1\}^*$ be a language. Two polynomial sized families of random variables $V$ and $S$ are \textit{statistically indistinguishable} for all poly-sized families of circuits $C$ when,
	\medskip
	$$|P[V, C, x] - P[S, C, x]| < |x|^{-c}$$
	for all constants $c>0$ and all sufficiently long $x \in L$.

\end{description}

%% file: chapters/methodologies/languages-with-zkps.tex
\section{Appropriate Problems for ZKP Systems}
We have explored what defines an interactive proof system and what makes it zero-knowledge, but what are concrete examples of ZKP systems, and which statements can we even prove in zero-knowledge?

Whether we can prove a statement in zero-knowledge depends on the underlying mathematical problem.
The problem also determines the ZKPs practical applications, simpler ZKPs are used to prove knowledge of a secret, while advanced ZKPs are used to prove signatures over committed values, set membership or range proofs \cite{camenisch2008efficient, bunz2018bulletproofs, camenisch2001efficient, bowe2018multi}.

The ZKP system \cite{goldwasser1989knowledge} used in our authentication system is based on the \textit{quadratic residuosity problem}.
We dive deep into how this ZKP system works, by exploring the mathematical foundation of quadratic residues, the quadratic residuosity problem, and the construction of the ZKP system.

We also look at examples of other problems and more broadly at classes of problems with ZKP systems.

\subsection{ZKP System for the Quadratic Residuosity Problem}
\label{zkp-qrp}
The first ZKP system defined \cite{goldwasser1989knowledge} is for the \textit{quadratic residuosity problem}.
The quadratic residuosity problem is much older than ZKPs. It was first described by Gauss in 1801 \cite{10.2307/j.ctt1cc2mnd}.
The problem emerges when computing quadratic residues with a modulo that is a product of two unknown prime numbers.
The properties of the problem enable it to be used as a \textit{trapdoor} function.
To define the problem, we must define the concept of quadratic residues and prime factorization.

\subsubsection{Quadratic Residues} 
The concept \cite{andrews1994number} comes from modular arithmetic.

\begin{remark}[Quadratic residues]
	For $x, n \in \mathbb{Z}$, $n > 0$, $\gcd(x, n) = 1 $.
	$x$ is a \textit{quadratic residue} if  $\exists w:w^2 \equiv x \Mod{n}$, otherwise $x$ is a \textit{quadratic non-residue}.
\end{remark}
\noindent For example,
$3$ is a quadratic residue mod 11, because $6^2 = 36 \equiv 3 \Mod{11}$.

\bigskip
\noindent
Generally, when $n$ is an odd prime, $x$ is a quadratic residue mod $n$ if,
$$x^{\frac{n-1}{2}} \equiv 1 \Mod{n}.$$

\paragraph{Legendre Symbol} $\dlegendre{x}{p}$ is a convenient notation for computation of quadratic residues, and is defined as a function of $x$ and $p$.
\bigskip
\newline
If $p$ is an odd prime then,
$$
\dlegendre{x}{p} =
		\begin{cases}
			1 & \text{$x$ is a quadratic residue modulo $p$}\\
			-1 & \text{$x$ is a quadratic non-residue modulo $p$}\\
			0 & \text{$gcd(x, p) \not = 1$}\\
		\end{cases}
$$
\smallskip
$$
\dlegendre{x}{p} \equiv x^{\frac{p - 1}{2}} \Mod{n} \quad \text{and} \quad \dlegendre{x}{p} \in \{-1, 0, 1\}
$$\\
Using the same example as before,

\begin{description}
	\item 3 is a quadratic residue modulo 11
	\medskip
	$$\dlegendre{3}{11}\equiv 3^{\frac{11-1}{2}} = 243 \equiv 1 \Mod{11}$$
	\item 6 is a quadratic non-residue modulo 11
	\medskip
	$$\dlegendre{6}{11}\equiv 6^{\frac{11-1}{2}} = 7776  \equiv -1 \Mod{11}$$

\end{description}

\paragraph{Jacobi Symbol}
A generalised definition of the Legendre symbol $\dlegendre{x}{n}$, to allow the case where $n$ is any odd number.

If $n = p_1p_2 \cdots p_r$, where $p_i$ are odd primes, then
$$\dlegendre{x}{n} = \dlegendre{x}{p_1}\dlegendre{x}{p_2} \cdots\dlegendre{x}{p_r}$$
\\
Unlike the Legendre symbol, if $\legendre{x}{n} = 1$, $x$ is a quadratic residue only if $x$ is a quadratic residue of every prime factor of $n=p_1p_2 \cdots p_r$.

\subsubsection{Prime Factorization}
The \textit{fundamental theorem of arithmetic} \cite{andrews1994number} states that for each integer \newline $n > 1$, exist primes $p_1 \le p_2 \le \cdots \le p_r$, such that $n = p_1 p_2 \cdots p_r$.

For example,

\begin{center}
	\begin{tabular}{|l|l|}
		\hline
		$1995 = 3 \cdot 5 \cdot 7 \cdot 19$ & 
		$1996 = 2^2 \cdot 499$ \\
		\hline
		$1997 = 1997$ &
		$1998 = 2 \cdot 3^3 \cdot 37$\\
		\hline
	\end{tabular}
\end{center}

Prime factorization is the decomposition of an integer $n$ to its prime factors $p_1 p_2 \cdots p_r$.
The problem is considered \textit{hard}, because currently no polynomial time algorithm exists for solving it \cite{Buchmann2001}. It is in class \textit{NP}, but is not proven to be \textit{NP-complete}.
The hardest instance of this problem is factoring the product of two prime numbers (\textit{semiprimes}).
The difficulty of this problem is a core building block in modern asymmetric cryptography like RSA \cite{rivest1978method}.

\subsubsection{Quadratic Residuosity Problem}

\begin{remark}[Quadratic Residuosity Problem]
	Given an integer $x$, a semiprime modulus $n = pq$, where $p$ and $q$ are unknown different primes, and a Jacobi symbol value $\legendre{x}{n} = 1$.
Determine if $x$ is a quadratic residue modulo $n$ or not.
\end{remark}

As mentioned before, the problem emerges when computing the quadratic residue with a modulo that is a product of two unknown primes.
The \textit{law of quadratic reciprocity} enables efficient computation of the Jacobi symbol value $\legendre{x}{n}$.
However, if $\legendre{x}{n} = 1$, it does not tell if $x$ is a quadratic residue modulo $n$ or not, $x$ is only a quadratic residue if $x$ is a quadratic residue of both modulo $p$ and $q$ ($\legendre{x}{p} = \legendre{x}{q} = 1$).
To calculate this, we would have to know the primes $p$ and $q$ by factoring $n$.
Since $n$ is a product of two prime numbers, factoring it is computationally hard.

The only efficient way to prove $x \Mod{n}$ is a quadratic residue, is with the root $w$.
The problem acts as a \textit{trapdoor} function, where it's hard to prove if $x \Mod{n}$ is a quadratic residue solely from $x$ and $n$, while it is easy to prove when you know $w$.

\subsubsection{Zero-Knowledge Proof Protocol}
To prove $x \Mod{n}$ is a quadratic residue in zero-knowledge we need to prove the existence of $w$, where $w^2 \equiv x \Mod{n}$, without revealing $w$ to the verifier.
Let us examine how the protocol \cite{goldwasser1989knowledge} defined by Goldwasser, Micali, and Rackoff achieves that.

\bigskip
\noindent
The Table \ref{table:zkp-qrp} contains a notation presenting an execution of the ZKP protocol for the quadratic residuosity problem.
The table displays consecutive steps in a process, the number on the left side of each row determines the step.
The columns under each party displays computations by that party, the column between parties displays the information exchanged and direction of the exchange.
This notation will be used in all further examples.

\bigskip
\begin{table}[h!]
	\caption{ZKP Protocol for the Quadratic Residuosity Problem}
	\centering
	\bigskip
	\begin{tabular}{rl}
		$n$ & Semiprime, where Jacobi $\legendre{x}{n} = 1$\\
 		$x$ & Residue, where $w^2 \equiv x \Mod n$\\
 		$w$ & Root\\
	\end{tabular}
	\medskip
	\begin{tabular}{r|r|c|l}
	\label{table:zkp-qrp}
		& Prover && Verifier\\
		\hline
		1&$u \leftarrow_R \Bbb{Z}_{n}^{*}; y = u^2 \Mod n$ & $\xrightarrow{y}$\\
		2 & & $\xleftarrow{b}$ & $b \leftarrow_R \{0, 1\} $\\
		3 &$z = uw^b\Mod n$ & $\xrightarrow z$ & verify $z^2 = yx^b \Mod n$\\
	\end{tabular}
\end{table}

\bigskip \noindent
The prover begins by picking up a random integer $u$ from the field $\Bbb{Z}_{n}$, computing $y = u^2 \Mod n$ and sending $y$ to the verifier.
The verifier picks a random bit $b$ and sends it to the prover. This random bit acts as the \textit{split in the tunnel} of our earlier cave analogy.
The prover computes the value $z$ based on $b$ and sends it over.
The verifier checks the proof by asserting $z^2 \equiv yx^b \Mod{n}$, this is possible since

$$z^2 \equiv yx^b \Mod{n}$$
$$(uw^b)^2 \equiv u^2(w^2)^b \Mod{n}$$
$$u^2w^{2b} \equiv u^2w^{2b} \Mod{n}.$$

\noindent For each round a cheating prover has a $\frac{1}{2}$ probability of succeeding by correctly guessing the value of the random bit $b$, to improve the confidence of the proof this is repeated $m$ times.

\subsection{Computational Complexity Classes}
We've looked at a ZKP protocol for a specific problem, but what other problems have ZKPs?
We can broadly examine the existence of ZKPs with classes of problems. This is a vast topic, so we will look only at some points. 
This knowledge is unnecessary for understanding the focus of our work, but offers an interesting background of ZKPs.

\paragraph{Non-deterministic Polynomial Time (NP).}
Class of problems solvable by a poly-time non-deterministic Turing machine, their proofs can be verified by a poly-time deterministic Turing machine.

Authors \cite{GMW} proved that every language in NP has a ZKP system, by defining a ZKP protocol for the Graph 3-Colouring problem (3-COL).
\textit{Minimum colouring problem} is a problem in graph theory, of what is the minimal $k$ \textit{proper} colouring of a graph, where no adjacent vertices are of the same colour.
An instance of ($k=3$) colouring (3-COL) is proven to be \textit{NP-Hard} because a polynomial reduction exists from \textit{Boolean-Satisfiability problem} (3-SAT) to 3-COL \cite{moore2011nature}.
According to Cook's theorem \cite{cook1971complexity} SAT or its 3 literal instance 3-SAT is \textit{NP-Complete}, and any language in $L \in NP$ can be reduced to an instance of 3-SAT. 
Furthermore because polynomial reductions are \textit{transient}, any language $L \in NP$ can be reduced to an instance of 3-COL.

%% file: chapters/results/intro.tex
\chapter{System Architecture and EAP Method Definition}
\thispagestyle{fancy}
\label{chapter:3}

\noindent 
In this chapter, we will focus on the architecture of our authentication system and the definition of the EAP method.
In the first section, we will examine how to utilise the ZKP protocol described in \S\ref{zkp-qrp} as a password verification method.
We need to be aware of the pitfalls of password authentication described in \S\ref{password-security-practices} and support necessary security practices.
In the second section, we will define our authentication system as an EAP method in the EAP authentication framework described in \S\ref{section:eap} by defining the messages and processes necessary for execution of our authentication system.

%% file: chapters/results/pba-using-qr-zkp.tex
\section{System Architecture}
\label{label:protocol-design}
In this section we will define the architecture of our authentication system, to do so we need to combine the model of password authentication we've examined in \S\ref{section:password-authentication} and the ZKP system for quadratic residuosity problem from \S\ref{zkp-qrp}.

\subsection{Password Verification}
\label{section:zkp-password-verification}
The purpose of password verification is to assert that the user authenticating knows the correct password $p$. 
How can we do this with the ZKP protocol?
The ZKP protocol proves that $x$ is a quadratic residue modulo $n$, by proving the knowledge of the root $w$, where  $w^2 \equiv x \Mod{n}$.

To use this protocol for password verification, we can use the password $p$ as the root $w$. The user and the authentication system can both follow the ZKP protocol, and the user will inevitably prove that he knows the password $w$, by proving that $x$ is a quadratic residue modulo $n$. With this, the system can assert that the user knows the correct password.

\subsubsection{Vulnerability}
To verify the proof provided by the user, the system needs to know the quadratic residue $x$.
Because the root $w = p$ is a password, this introduces a vulnerability as mentioned in \S\ref{label:password-vulnerabilities}.
An attacker with access to $x$ could crack the password $w$ in an offline attack with pre-computed tables.
As mentioned in \S\ref{paragraph:password-hashing}, we need to use a key-stretching method to ensure adequate security against such offline attacks.

\subsubsection{Theoretical Constraints of Key-Stretching Vulnerable Data}
A common usage of a key-stretching method is to transform the vulnerable data stored in the authentication system.
However, this approach doesn't work in our case.
Let us explore how the authentication system verifies the proof, and why using key-stretching directly over stored data is an issue.

We assume the system can verify the proof and use key-stretching methods directly over the vulnerable data. 
However, we will see why this is not possible.

\paragraph{Proof Verification with Key-Stretched Data}
\label{paragraph:problems-with-key-stretch}
On the last step of the protocol the system verifies that
$$ z^2 \equiv yx^b \Mod{n}.$$
If we stretch the vulnerable value $x$ with a function $H$ and a salt $s$
$$H(x, s) = x_H,$$
we can then verify the proof with an inverse function $H^{-1}$
$$z^2 = yH^{-1}(x_H, s)^b.$$
This is possible assuming a polynomial algorithm $H^{-1}$ exists, however, since key-stretching methods are based on hashing functions which are one-way functions, we know that the probability of a polynomial algorithm $H^{-1}$ to successfully compute a \textit{pseudo-inverse} is negligibly small, for all positive integers $c$ \cite{goldreich2007foundations}
$$\Pr[H(H^{-1}(H(x))) = H(x)] < |x|^{-c}.$$
Even if given unbounded time and resources, the \textit{pseudo-inverse} $x' = H^{-1}(H(x))$ might not be equal to $x' \not = x$. 
The set $x'\in I_x$ are all values that map into $H(x) = H(x')$, and since $H$ is not injective we know that $|I_x| \ge 1$.
Meaning that the probability that $x' = x$ is
\medskip
$$\Pr[H^{-1}(H(x)) = x] = \frac{1}{|I_x|}.$$

\subsubsection{Key-Stretching the Root $w$}
We've seen that key-stretching the vulnerable value $x$ prevents us from verifying the ZKP.
However, by increasing the entropy of the root $w$, we can eliminate the vulnerability and ensure adequate security.

\bigskip
\noindent Instead of treating the password $p$ as the root $w$, we can instead derive the root $w$ from the password $p$, by stretching it with a function $H$ and a salt $s$
$$w = H(p, s),$$
and use the output as the root $w$.
This way we've ensured the same level of protection as if we stretched the data stored in the system.
Because we didn't change the value $x$, we can verify the proof without being affected by issues mentioned in \S\ref{paragraph:problems-with-key-stretch}.

\subsection{Secure Authentication Process using ZKPs}
By key-stretching the password to derive the root $w$, we've figured out how to secure our system while respecting the constraints imposed by the proof verification process.
How does this change the authentication process we've described in \S\ref{section:zkp-password-verification}?

\bigskip
\noindent
In Table \ref{table:sap-zkp}, variables in with an $_i$ subscript (e.g. $y_i$) are unique to each iteration $i$.

\begin{table}[h!]
	\centering
		\caption{Secure Authentication Process using ZKPs}
		\vspace{0.1cm}
	\begin{tabular}{p{0.016\textwidth}|p{0.25\textwidth}|p{0.03\textwidth}|p{0.3\textwidth}}
  		& User & & Authentication System\\
  		\hline
		1 & $w = H(p, s)$ & & \\ 
		\hline
		2 & $u_i \leftarrow_R \Bbb{Z}_{n}$ &  \\
		& $y_i = u_i^2$ & $\xrightarrow{y_i}$ \\
		3 & & $\xleftarrow{b_i}$ & $b_i \leftarrow_R \{0, 1\} $ \\
		4 & $z_i = u_iw^{b_i} \Mod n $ & $\xrightarrow{z_i}$ & check $z_i^2 \equiv y_ix^{b_i} \Mod{n}$\\ 
	\end{tabular}
	\label{table:sap-zkp}
\end{table}

\bigskip
\noindent
The process (Table \ref{table:sap-zkp}) will now begin with the user computing the root $w$ from the password $p$ and salt $s$.
Once the user computes the root $w$, he can authenticate by following ZKP protocol with the system, as mentioned in \S\ref{zkp-qrp}
Earlier we argued the ZKP works as a password verification method because $p = w$, this argument isn't true anymore.
However, even though $w \not = p$, the user can only derive $w$ knowing the password $p$, so when the user proves the knowledge of $w$, it can only be so because they know $p$ as well.

\noindent
We repeat this part of the process $m$ times for a confidence of $1 - 2^{-m}$.
 
\subsection{Considerations}
\label{section:pefromance-considerations}
\paragraph{Performance.}
If we look at the steps that occur in our ZKP verification process (Table \ref{table:sap-zkp}), we can notice iterations of data exchanges between the user and the system.
In a real world environment, this can cause the authentication process to be slow because of network inefficiencies when transmitting data between the user and the authentication system.
However since iterations are mutually independent, we can execute them in parallel (Table \ref{table:zkp-qrp-parallel}).

\begin{table*}[h!]
	\centering
		\caption{Parallel ZKP Construction}
		\vspace{0.2cm}
	\resizebox{13cm}{!}{
	\begin{tabular}{l|l|c|l}
  		& Peer & & Authenticator\\
  		\hline
		1 & $w = H(p, s)$ & & \\ 
		\hline
		2 & $u_1,...,u_m \leftarrow_R \Bbb{Z}_{n}$ &  \\
		& $y_1,...,y_m = u_1^2,...,u_m^2 \Mod{n}$ & $\xrightarrow{y_1,...,y_m}$ \\
		3 & & $\xleftarrow{b_1,...,b_m}$ & $b_1,...,b_m \leftarrow_R \{0, 1\} $ \\
		4 & $z_1,...,z_m = u_1w^{b_1},...,u_mw^{b_m} \Mod n $ & $\xrightarrow{z_1,...,z_m}$ & check $z_1^2, ..., z_m^2 \equiv y_1x^{b_1},...,y_mx^{b_m} \Mod{n}$\\
	\end{tabular}
	}
	\label{table:zkp-qrp-parallel}
\end{table*}

What we are proposing is theoretically called a 3-round interactive ZKP.
The existence of these proofs is limited only to a class of problems \textit{BPP} \cite{goldreich1996composition}.
Unfortunately, the quadratic residuosity problem is not believed to be in this class, so we assume a parallel proof to have a weaker notion of zero-knowledge.

For this purpose, we've used a sequential execution for our authentication process.

%% file: chapters/results/eap_pzkpqr.tex
\section{EAP Method Definition}
\label{section:eap-84-definition}
We want to encapsulate our extended zero-knowledge authentication system defined in \S\ref{label:protocol-design} as an EAP method in the EAP framework we've explored in \S\ref{section:eap}.
To achieve this, we must define a new EAP method, which comprises of the messages exchanged between the \textit{peer} and the \textit{authenticator}, their data formats and the processes for handling them.

\paragraph{Terminology.}
In this section we will use EAP terminology as described in \S\ref{section:eap}, which uses different names to describe parties involved, as the ones used in our system architecture in \S\ref{label:protocol-design} or as in the ZKP protocol in \S\ref{zkp-qrp}.
The two parties in EAP are called the peer and the authenticator, where the peer is authenticating with the authenticator.
To draw parallels between our system architecture, where we use the names \textit{user} and \textit{authentication system}, the peer is the user, and the authenticator is the authentication system.
In the ZKP protocol names \textit{prover} and \textit{verifier} are used, the peer is the prover, and the authenticator is the verifier.

\bigskip
\noindent
To define an EAP method, we need to break down our authentication system described in \S\ref{label:protocol-design} to EAP messages representing interactions between the user and the authentication system.
Each message defines its data format, the sender and recipient processes and local state changes.
Our EAP method defines two messages, the \textit{setup phase} message and the \textit{verification phase} message.

We designed our authentication system for multiple users. 
For this reason, the authenticator needs to start the authentication process with the \textit{identification phase} by querying the identity of the peer with the \textit{identity} (Type 1) EAP message as described in \S\ref{def:eap-identitiy}.
In the \textit{setup phase} the peer uses the \textit{setup} message for discovery of ZKP parameters, and to provide the values for the first proof verification round to the authenticator.
This message is exchanged only once.
In the \textit{verification phase} the \textit{verification} message is used to exchange data required for a single round of proof verification. Since proof verification as described in \S\ref{zkp-qrp} requires $m$ iterations, this message is exchanged $m$ times.
The protocol ends with the authenticator sending either a \textit{success} or a \textit{failure} message.

\subsubsection{Authentication process overview}
\begin{figure}[h!]
	\centering
	\includegraphics[width=14cm]{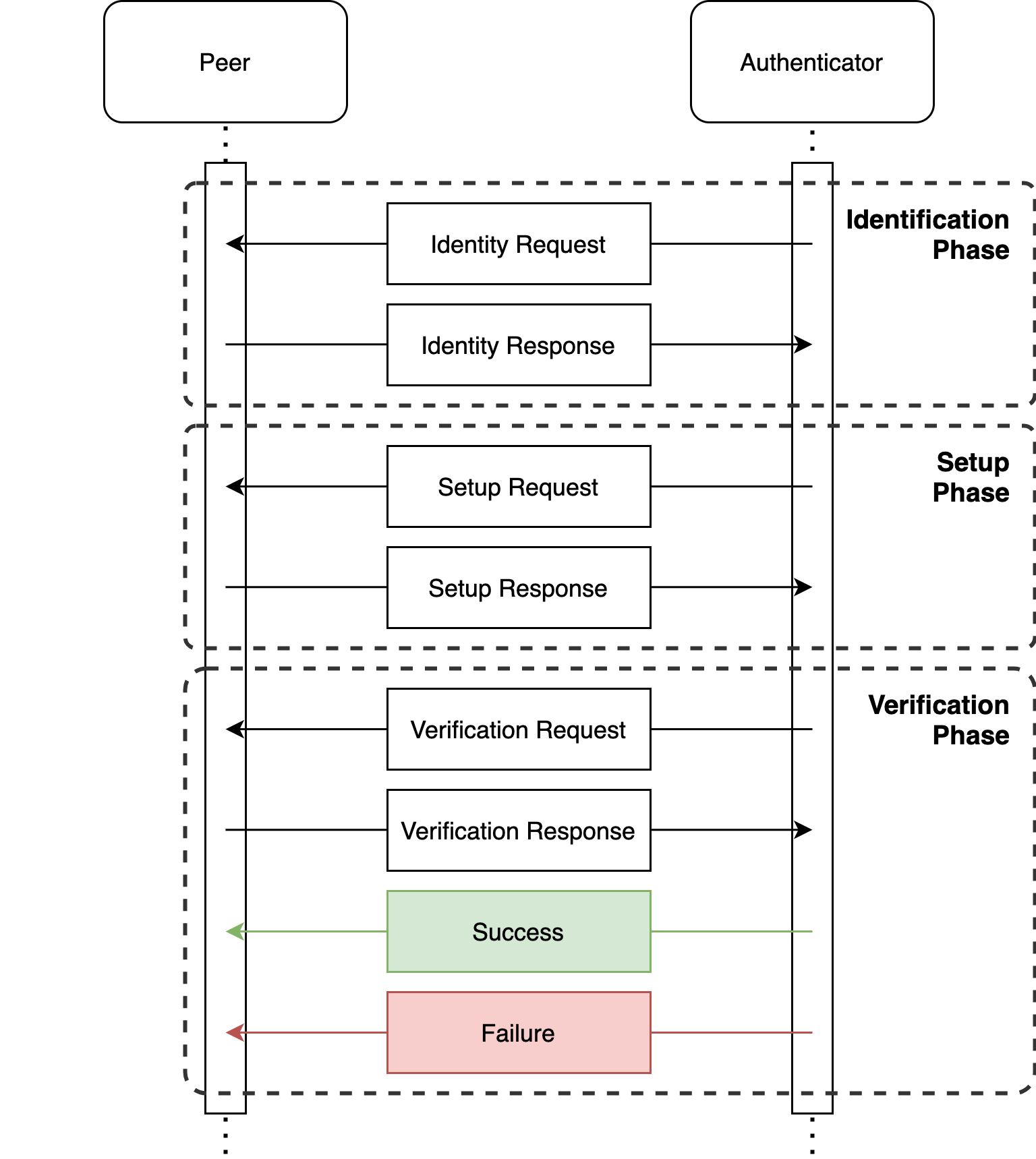}
	\caption{EAP Method Execution}
	\label{fig:eap-84}
\end{figure}

Let us examine the EAP messages (Figure \ref{fig:eap-84}) of the authentication process described in Table \ref{table:zkp-qrp-2}.
The mapping between EAP messages and the steps in Table \ref{table:zkp-qrp-2} is \textit{not one-to-one}, as we merged some steps to reduce the number of message exchanges to speed up the whole process.
In this section we present a simplified authentication process to present the general idea, while a detailed description is given in the next section.

The mathematical variables have the same meaning as in the system architecture described in \S\ref{label:protocol-design}. This applies to all further sections.
\begin{table}[h!]
	\centering
	\caption{Improved ZKP Authentication with EAP}
	\vspace{0.2cm}
	\begin{tabular}{l|l|c|l}
  		& Peer & & Authenticator\\
  		\hline
  		1 & & $\xrightarrow{I}$ &\\
		2 & $w = H(p, s)$ & $\xleftarrow{s,n}$ & \\
		\hline
		3 & $u_i \leftarrow_R \Bbb{Z}_{n}$ &  \\
		& $y_i = u_i^2 \Mod{n}$ & $\xrightarrow{y_i}$ \\
		4 & & $\xleftarrow{b_i}$ & $b_i \leftarrow_R \{0, 1\} $ \\
		5 & $z_i = u_iw^{b_i} \Mod n $ & $\xrightarrow{z_i}$ & check $z_i^2 \equiv y_ix^{b_i} \Mod{n}$\\
	\end{tabular}
	\label{table:zkp-qrp-2}
\end{table}

The authentication process (Figure \ref{fig:eap-84}) consists of multiple phases:

\begin{description}
	\item [Identification Phase] Used to establish the identity of the peer. The authenticator sends an \textit{identity request} message, and the peer responds with an identifier $I$, which is used the locate the salt $s$ and quadratic residue value $x$. (Table \ref{table:zkp-qrp-2}, Step 1.)
	\item [Setup Phase] Used to exchange necessary values for the \textit{verification phase}. The authenticator sends the salt $s$ and modulus $n$ in the \textit{setup request} message, which the peer uses to derive the root $w$ from the password $p$. The peer responds with $y_1$ which will be used in the first round of the verification phase. (Table \ref{table:zkp-qrp-2}, Steps 2. and 3.)
	\item [Verification Phase] In this phase both parties continuously exchange data for $m$ rounds to construct and verify the proof. In a given round $i$ the authenticator sends the random bit $b_i$ and the peer responds with the partial proof $z_i$. The peer also sends the $y_{i+1}$ for the next verification round $i+1$, this is done as a performance optimization (Table \ref{table:zkp-qrp-2}, Steps 4., 5. and 3. again). After receiving each response the authenticator verifies the partial proof, once he has done so for $m$ successfully, he sends a \textit{success} message. If the partial proof isn't valid, the authenticator must send a \textit{failure} message.
\end{description}

\subsubsection{Message Format}

\begin{center}
\begin{tabular}{|c|c|c|c|}
	\hline
	Length (Octets) & 1 & $k$\\
	\hline
	Field & Phase Type & Phase Data\\
	\hline
\end{tabular}
\end{center}

\noindent
An EAP message is composed of many fields (Table \ref{table:eap-message}).
The \textit{type} field as described in \S\ref{text:eap:type} determines the EAP method used by the peer and the authenticator to interpret the \textit{type-data} field.
Our EAP method is indicated with \textit{type} value $84$.

The \textit{type-data} field is composed of two sub fields, a \textit{phase type} field and a \textit{phase data} field.
\textit{Phase} identifies the phase of the authentication process, while \textit{phase data} holds specific data for that phase.
We have defined 2 \textit{phase} values, that correspond with the phases of our authentication process:

\begin{quote}
	\begin{description}
		\item [Phase Type 1]: Setup Message (Figure \ref{fig:eap-84} - Setup Phase)
		\item [Phase Type 2]: Verification Message (Figure \ref{fig:eap-84} - Verification Phase)
	\end{description}
\end{quote}

The \textit{Identification Phase} in Figure \ref{fig:eap-84} doesn't have a corresponding \textit{phase type} value as the \textit{Identity} message as described in \S\ref{def:eap-identitiy} is a pre-defined EAP message type.

Note also, that the architecture does not specify a key-stretching method, and neither does this EAP method.
We assume that in a practical implementation the method would be pre-defined and implicitly known by both parties.

\subsection{Setup Phase}
Previously we assumed the modulus $n$ and salt $s$ are known by the user, however, the EAP method needs to facilitate the discovery of this data.
Our system is designed to support multiple users, so before this phase the authenticator needs to identify the peer with the \textit{identity} message.
The response message additionally contains the control value $y_1$, used to verify the proof in the first verification round. The data is already included in this message to improve the method performance.

\subsection*{Request Message} Message is used to deliver the salt $s$ and semiprime modulus $n$ to the peer.

\paragraph{Phase Data Format}

\begin{center}
\begin{tabular}{|c|c|c|c|}
	\hline
	Length (Octets) & 1 & $4 \le k \le 255 $ & $64 \le j$\\
	\hline
	Field & Salt Length & Salt & Modulus\\
	\hline
\end{tabular}
\end{center}

\begin{quote}
\begin{description}
	\item[Salt Length] A single octet for the length of the \textit salt field in octets.
	\item[Salt] A random salt value, should be from 4 octets to 255 octets long.
The max length is determined by the largest number able to be encoded in the \textit {salt length} field.
	\item[Modulus] Fills the rest of the message to the length specified by the \textit{length} field in the EAP message. 
Should be at least 64 octets (512 bits).
\end{description}
\end{quote}

\paragraph{Request Handling} When a request is received, the peer computes the root $w$ using the password $p$ the salt $s$ with the pre-determined hashing function $H$.
Root $w$ should be stored stored in memory by the peer. 
Next the peer should pick a random integer $u$ from field $Z^*_n$, and store it in memory and then computes the control value $y = u^2 \Mod{n}$.
The control value $y$ is included in the response message.

\subsection*{Response Message}
The response contains the control value $y_1$ to the authenticator to be used in the first round of the verification process.
The subscript of a variable $y_i$ denotes in which verification round $i$ is the variable used.

\paragraph{Phase Data Format}

\begin{center}
\begin{tabular}{|c|c|}
	\hline
	Length (Octets) & $k$ \\
	\hline
	Field & Control Value $y_1$\\
	\hline
\end{tabular}
\end{center}

\bigskip
\begin{quote}
\begin{description}
	\item[Control Value] Computed by the peer, where $y_1 = u_1^2 \Mod{n}$ and $u_1 \leftarrow_R \mathbb{Z}^*_n$.
\end{description}
\end{quote}

\paragraph{Response Handling}
The authenticator should store the $y_1$ control value locally to be used when verifying the proof $z_1$.

\subsection{Verification Phase}
This message pair exchanges data required to compute and verify the proof.
They continuously exchange it until the authenticator concludes the authentication.
After $m$ successful rounds, when the authenticator reaches a confidence of $1 - 2^{-m}$ in the proof, the authentication is successful.
To make our method more efficient, we reduce the number of exchanged messages between the parties by interlacing some data between iterations.
On the round $i$, the response contains data required for the round $i+1$.

\subsection*{Request Message}
In the round $i$, the authenticator generates random bit $b_i$ stores it locally, and sends it to the peer.
\paragraph{Message Data Format}

\begin{center}
\begin{tabular}{|c|c|}
	\hline
	Length (Octets)  & $1$ \\
	\hline
	Field & Random Bit $b_i$\\
	\hline
\end{tabular}
\end{center}

\begin{quote}
\begin{description}
	\item[Random Bit] A single-bit $b_i$, at the right-most place. 1 octet long.
\end{description}
\end{quote}

\paragraph{Request Handling}
The peer computes the proof $z_i = u_iw^{b_i} \Mod{n}$, with the bit $b_i$ received in the request.

Additionally the peer generates the control value $y_{i+1}$ for the next ($i+1$) verification round.
The peer picks a random integer $u_{i+1}$ from field $Z^*_n$ and stores it in memory.
The control value is computed as $y_{i+1} = u_{i+1}^2 \Mod{n}$ and sent in the response.

\subsection*{Response Message}
The response transmits the proof $z_i$ and the control value $y_{i+1}$ to the authenticator, who verifies the proof and decides on how to proceed.

\paragraph{Message Data Format}

\begin{center}
\begin{tabular}{|c|c|c|c|}
	\hline
	Length (Octets) & $1$ & $k $ & $j$\\
	\hline
	Field & Proof Length & Proof $z_i$ & Control Value $y_{i+1}$\\
	\hline
\end{tabular}
\end{center}

\begin{quote}
\begin{description}
	\item [Proof Length] A field one octet in length. Determines the length of the Proof field in octets.
	\item [Proof] Value $z_i$ computed by the peer, verified by the authenticator.
	\item [Control Value] Value $y_{i+1}$, required to verify the proof of the $(i+1)$-th round.
\end{description}
\end{quote}

\paragraph{Response Handling}
The authenticator should verify the proof by asserting that $z_i^2 \equiv y_ix^{b_i} \Mod{n}$.
If the verification fails, the a \textit{failure} message must be sent to the peer, otherwise a \textit{success} message must be sent if the verification was successful for $m$ rounds.
If that is not the case, the $y_{i+1}$ is stored by the authenticator and a new verification message request is sent.

%% file: chapters/results/conclusion.tex
\chapter{Conclusion and Future Work}
\thispagestyle{fancy}
\label{chapter:4}
The aim of this thesis was to study the use of ZKPs as an authentication mechanism.
In section \S\ref{label:protocol-design}, we have presented the architecture of an authentication system, which uses a ZKP protocol as the password verification method.
We have described how the ZKP protocol can prove the knowledge of the user's password. The architecture also supports key-stretching for protection against password vulnerabilities discussed in \S\ref{label:password-vulnerabilities}.
In section \S\ref{section:eap-84-definition}, we have encapsulated our authentication system within EAP by defining a specification for a new EAP method.
The specification contains definitions for EAP messages and their handling procedures.

We have been successful in our goal of studying and using the ZKP protocol.
However, upon observation, the system performance is not on par with today's industry standards. The iterative nature of the underlying ZKP protocol accumulates communication latencies, slowing down the system.

\paragraph{Future work.}

\begin{itemize}
	\item The EAP method specification presented in this work can be implemented and tested in a real-world environment.
	\item The ZKP protocol used in this work is a first generation protocol. Today there are many newer protocols that have solved many shortcomings of the older generation ZKPs. Using a newer generation ZKP protocol can improve the performance of the authentication system.
	\item The ZKP protocol we've examined is iterative, which can cause worse performance. We've discussed an alternative proof construction in \S\ref{section:pefromance-considerations}, which we aren't pursuing in this thesis because of assumed weaker strength of zero-knowledge. However, in a real-world application, the performance improvements might justify the theoretical shortcomings.
\end{itemize}

%% file: chapters/slo-summary.tex
\chapter{Povzetek naloge v slovenskem jeziku}
\thispagestyle{fancy}
Problematika zasebnosti je vsak dan večja zaradi vse večje prisotnosti informacijskih sistemov v naših življenjih. Zdi se, da se zasebnost in tehnologija medsebojno izključujeta. Dokazi ničelnega znanja (ZKP) imajo potencial, da spremenijo, kako naši osebni podatki obstajajo v digitalnem prostoru.
V zaključni nalogi raziščemo preprosto uporabo ZKP kot metodo preverjanja gesla v avtentikacijskem sistemu.

V delu najprej predstavimo področje avtentikacije z avtentikacijskimi sistemi ter mehanizmom avtentikacije z gesli, njihovimi ranljivostmi in tehnikami za zaščito pred njimi.
Spoznamo tudi razširljivo avtentikacijsko ogrodje (EAP), v katerega je sistem umeščen.
Končno predstavimo tudi dokaze ničelnega znanja (ZKP), ki služijo kot naša metoda preverjanja veljavnosti gesla.

Avtentikacija je v splošnem proces preverjanja resničnosti nekih trditev. \linebreak V računalništvu je zelo pogosta oblika le-tega uporaba uporabniškega imena in gesla za vzpostavljanje dostopa med uporabnikom in sistemom.
Varnostni model tega sistema temelji na principu deljene skrivnosti med sistemom in uporabnikom.
V postopku avtentikacije z gesli, se uporabnik najprej identificira v sistemu z uporabniškim imenom, nato pa s sistemom deli geslo, ki ga sistem primerja z lokalno shranjenimi podatki.
Sistem je zelo preprost, kar ga naredi ranljivega za napade brez povezave, zato varne implementacije takšnih sistemov uporabljajo metodo raztegovanja ključev, da se pred tem zaščitijo.

Naš avtentikacijski sistem bo umeščen kot metoda v razširljivo avtentikacijsko ogrodje (EAP).
EAP je splošno namensko avtentikacijsko ogrodje, zasnovano za omrežno avtentikacijo. EAP definira knjižnico postopkov, metod in sporočil, prek katerih se lahko avtentikacijski sistem in vrstnik uskladita in izvršita množico avtentikacijskih protokolov. 
EAP je pogosto uporabljen kot avtentikacijski sistem za brezžična omrežja.

Kot mehanizem preverjanja gesla uporabljamo dokaze ničelnega znanja (ZKP).
ZKP je način dokazovanja matematičnih trditev, ki lahko dokaže, da je trditev resnična, brez da bi razkrili, zakaj je trditev resnična.
Za razliko od matematičnih izrekov je ZKP verjetnostni, ker preverjevalca prepriča, da je trditev resnična z zanemarljivo majhno verjetnostjo napake.
ZKP-ji so popularno orodje v kriptovalutah, kot so Zcash, Monero, Ethereum ter Solana. Uporabljajo se tudi v sistemih digitalne identitete, kot je Idemix.

Interaktivni sistemi dokazovanja so teoretično ogrodje, v katerem so definirani interaktivni ZKP-ji. V takem sistemu poskuša dokazovalec prepričati preverjevalca v resničnost neke trditve. 
V takem sistemu velja, da iskreni dokazovalec lahko prepriča preverjevalca v resničnost neke trditve ter da goljufivi dokazovalec ne bo nikoli prepričal preverjevalca, ki pravilno sledi protokolu.

ZKP-ji dokažejo resničnost trditve, brez da bi razkrili, zakaj je resnična. 
Na primer da sta dve žogi različne barve, brez da razkrijemo barve same.
Osrednja ideja za ničelnim znanjem je, da zunanji opazovalec ne more ločiti med podatki, ki so bili izmenjani med izvajanjem ZKP protokola, in podatki, ki oponašajo izvajanje ZKP protokola.
Takšne podatke lahko ustvari kdorkoli, zato lahko sklepamo, da dokler so “pravi” podatki nerazločljivi, iz njih ne moremo izčrpati nobenega novega “znanja”.

Zmožnost dokazovanja neke trditve z dokazi ničelnega znanja je odvisna od \linebreak matematičnega problema, za katerega trditev obstaja.
Vrsta problema določa tudi način uporabe ZKP-ja. S preprostimi protokoli lahko dokažemo na primer poznavanje skrivnega ključa. 
Z naprednimi ZKP-ji lahko dokažemo skorajda karkoli, kar je mogoče preveriti v poljubnem algoritmu.
Naš mehanizem preverjanja gesla uporablja preprosti ZKP protokol, osnovan na problemu kvadratnih ostankov.
Problem kvadratnih ostankov se pojavi v modularni aritmetiki pri računanju kvadratnih ostankov, kjer je modulo zmnožek dveh neznanih praštevil.
Problem kvadratnih ostankov je težak, ker se v njem skriva problem razčlenjevanja praštevil, za katerega učinkoviti algoritmi ne obstajajo.
Lastnosti tega problema ga naredijo primernega za funkcijo "zapornih vrat", kjer je operacija v eno stran lahka, v nasprotno stran pa zelo težka, če ključa ne poznamo. 
Da je neko število kvadratni ostanek modulo n, lahko učinkovito dokažemo z obstojem korena, iz katerega ostanek izhaja. 
ZKP protokol dokaže, da je neko število kvadratni ostanek modulo n tako, da dokaže obstoj korena, brez da bi razkril koren sam.


V arhitekturi avtentikacijskega sistema smo združili mehanizem varne avtentikacije z gesli in ZKP-jem za problem kvadratnih ostankov.
Spomnimo se, da v procesu avtentikacije z gesli uporabnik dokaže, da pozna geslo tako, da ga pošlje preko omrežja in sistem preveri, če se ujema z lokalnimi podatki.
Če smatramo geslo kot koren v ZKP protokolu, lahko postopek preverjanja gesla nadomestimo z ZKP protokolom.
Takoj ko uporabnik dokaže obstoj korena, dokaže tudi, da pozna geslo.
V arhitekturo moremo vključiti tudi metodo raztegovanja ključev, saj je neposredna uporaba gesla ranljiva za napade brez povezave.

%
Avtentikacijski sistem, umeščen znotraj EAP ogrodja, kot nova EAP metoda.
Metoda se izvaja v treh fazah; fazi identifikacije, nastavitve in preverjanja.
V fazi identifikacije se vrstnik identificira sistemu, ta faza uporablja obstoječ tip sporočila z EAP ogrodjem.
V naslednji fazi nastavitve, si vrstnik in sistem izmenjata parametre za izvršitev ZKP protokola in metode raztegovanja ključev.
V zadnji fazi si vrstnik in avtentikacijski sistem izmenično pošiljata izzive in dokaze. Po m uspešnih ponovitvah sistem uspešno avtenticira vrstnika. 
Če se pri kateremkoli koraku zalomi, je proces prekinjen.

%

Avtentikacijski sistem deluje, vendar je še veliko stvari, ki bi jih lahko izboljšali.
Iterativna izvedba protokola ni idealna za uporabo prek omrežja, kjer čas pošiljanja podatkov zelo poslabša čas avtentikacije.
Postopek lahko pohitrimo tako, da zaporedne korake postopka izvajamo vzporedno. S tem skrajšamo čas avtentikacije na konstantno dolžino.
Uporabljen ZKP protokol je eden prvih, ki je bil zasnovan, zato bi bilo mogoče bolje, če bi uporabili moderen protokol.